\pdfoutput=1

\documentclass[nonacm,sigconf]{acmart}
\renewcommand\footnotetextcopyrightpermission[1]{} 

\AtBeginDocument{%
  \providecommand\BibTeX{{%
    \normalfont B\kern-0.5em{\scshape i\kern-0.25em b}\kern-0.8em\TeX}}}

\usepackage{multirow,tabularx}
\usepackage{makecell}

\usepackage{float}
\usepackage{subfig}

\usepackage[font=small]{caption}

\usepackage{fontawesome5}

\begin{document}

\title{Rating and aspect-based opinion graph embeddings for explainable recommendations}

\author{Iván Cantador}
 \affiliation{%
   \institution{Universidad Autónoma de Madrid}
   \city{Madrid}
   \state{Spain}
 }

\author{Andrés Carvallo}
 \affiliation{%
   \institution{Pontificia Universidad Católica de Chile}
   \city{Santiago}
   \state{Chile}
 }

\author{Fernando Diez}
 \affiliation{%
   \institution{Universidad Autónoma de Madrid}
   \city{Madrid}
   \state{Spain}
 }


\begin{abstract}
The success of neural network embeddings has entailed a renewed interest in using knowledge graphs for a wide variety of machine learning and information retrieval tasks. In particular, recent recommendation methods based on graph embeddings have shown state-of-the-art performance. In general, these methods encode latent rating patterns and content features. Differently from previous work, in this paper, we propose to exploit embeddings extracted from graphs that combine information from ratings and aspect-based opinions expressed in textual reviews. We then adapt and evaluate state-of-the-art graph embedding techniques over graphs generated from Amazon and Yelp reviews on six domains, outperforming baseline recommenders. Additionally, our method has the advantage of providing explanations that involve the coverage of aspect-based opinions given by users about recommended items.
\end{abstract}

\ccsdesc[500]{Information systems~Recommender systems}
\ccsdesc[500]{Computing methodologies~Semantic networks}
\ccsdesc[500]{Computing methodologies~Neural networks}

\keywords{recommender systems, knowledge graphs, graph embeddings}


\maketitle

\vspace{-1mm}

\section{Introduction}
In recent years, knowledge graphs (KG) have been widely used in recommendation~\cite{he2015trirank, wang2018ripplenet}. A KG is a graph data structure containing information about semantic entities (or concepts) expressed as nodes, and semantic relations between entities expressed as edges, and the relation can be represented as \textit{<subject, property, object>} triples whose elements may belong to structured knowledge bases or ontologies. Subject and object entities (nodes) can refer to users, items, item metadata, and content features, among others.
Compared to recommendation approaches that treat relations between users and items independently, exploiting a KG graph allows capturing deeper relations valuable to generate better recommendations~\cite{wang2019kgat}. Another benefit of a KG-based recommender is its capacity to deal with cold-start situations for new users or unrated items by providing side information~\cite{hu2018leveraging, huang2018improving}. Moreover, it allows generating human-interpretable recommendations, making the system more reliable and transparent for the end-user ~\cite{wang2019explainable, he2015trirank, xian2019reinforcement}.

The broad majority of graph-based recommendation methods makes use of information underlying rating relations and connections between items with similar attributes~\cite{wang2019explainable}. In general, these methods encode latent rating patterns and content features using graph-embeddings, since the information stored in KG triples can be too computationally expensive to manipulate. Knowledge-graph-embeddings~\cite{cai2018comprehensive} consist of learned lower dimension, continuous vectors for entities and relations that preserve the original structure of the KG. 

In this context, a source of information that has been barely explored for KG is the textual content of user reviews. In the research literature, it has been shown that both ratings and reviews can give support and trust to users in decision making tasks~\cite{lackermair2013importance}. In particular, approaches have been proposed to automatically extract personal opinions about item aspects (i.e., features, components, and functionalities) from reviews~\cite{poria2016aspect}. Aspect-based opinions have been used to enrich sentiment analysis models~\cite{de2018aggregated}, document classification~\cite{ostendorff2020aspect}, and recommender systems~\cite{he2015trirank}. Concerning the exploitation of aspect-based opinions by recommender systems, Musto et al.~\cite{musto2019combining} made use of aspects to justify recommendations using sentiment analysis and text summarization, whereas Wu and Ester~\cite{wu2015flame} proposed a probabilistic approach that combines aspect-based opinion mining and collaborative filtering. Aspect-based opinions have also been included as additional features for latent factor~\cite{qiu2016aspect} and neural network~\cite{chin2018anr, guan2019attentive} models. To the best of our knowledge, the approach by He et al.~\cite{he2015trirank} is the only one that takes advantage of a KG with information from aspect-based opinions extracted from reviews. Differently, our method combines rating and aspect-based opinion interactions in a KG embedding framework to obtain a richer representation of users, items, and aspects in the same latent space for making recommendations.

Therefore, the main contributions from our work are as follows:

\begin{enumerate}
\item Presenting a graph model that represents and relates users, items, and aspects in the same latent space for its exploitation by recommender systems.
\item Using the graph model to adapt recommendation methods based on neural network graph embeddings, showing empirical performance improvements over state-of-the-art baselines. 
\item Proposing a method that uses the graph model to provide explanations of generated recommendations at the aspect opinion level.
\end{enumerate}


\section{Related work}

\textbf{Graph-based recommendations.} We can distinguish between three main types of approaches that use graphs for recommendation purposes: path-based, embedding-based and unified. 
In \textit{path-based methods} take advantage of users and items connectivity score functions to generate recommendations. For instance, Luo et al.~\cite{luo2014hete} proposed Hete-CF, which makes use of path similarity combined with common paths between two entities in a social network to make recommendations. Concerning \textit{embedding-based methods}, the KG is used to obtain latent representations of users and items, with which finding closest users and items. Examples of this are the work by Wang et al.~\cite{wang2018dkn} for news recommendation and Huang et al.~\cite{huang2018improving}, which incorporates sequential information through a recurrent neural network. Another recent work based on embeddings is KGAT~\cite{wang2019kgat}, which models high order relations in a KG. Finally, \textit{unified methods} exploit information from embedding- and path-based forms together. Wang et al.~\cite{wang2018ripplenet} proposed RippleNet, which introduces the concept of preference propagation to refine user and item representations obtained with embedding-based methods. In our work, we advocate for embedding-based graph recommendations, since they allow better addressing the cold start problem and efficiently representing entities and relations, preserving the graph structure~\cite{wang2019kgat, wang2018ripplenet}. Moreover, we chose the embedding-based approach since the state-of-the-art KGAT method~\cite{wang2019kgat} achieved better performance results than other embedding-based, path-based and unified approaches. This led us to empirically compare our method with KGAT in the experimental section.

\textbf{Aspect-based recommendations.} Many aspect-based recommenders exist in the research literature, but few consider the incorporation of opinions about item aspects into a KG. Aspect-based opinions have been used to justify recommendations, including sentiment analysis and text summarization from reviews~\cite{musto2019combining}. The combination of aspect-based opinion mining and collaborative filtering in a probabilistic framework was proposed in~\cite{wu2015flame}. Moreover, using aspects as features of latent factor~\cite{qiu2016aspect} and neural network~\cite{chin2018anr, guan2019attentive} models have also been investigated. Regarding the incorporation of aspect-based opinion information into a KG, TriRank~\cite{he2015trirank} makes use of a tripartite graph representing relationships between users, items, and aspects to generate recommendations. Differently to TriRank, our method combines ratings and aspect-based opinions in a KG to learn richer vector representations for both users and items, with which generating personalized recommendations. Furthermore, there have been approaches to automatize text content analysis in the field of medicine\cite{carvallo2020automatic, carvallo2019comparing}, along with stress tests\cite{aspillaga2020stress, araujo2021stress, araujo2020adversarial} to evaluate the models performance under different situations.

\textbf{Graph-based recommendation explanations.} The majority of works on graph-based recommenders aim to produce path-based interpretations of generated recommendations, although it has been stated that they require domain knowledge~\cite{ma2019jointly}. Fu et al.~\cite{fu2020fairness} made use of paths from users to items, mitigating unfairness problems related to undesired recommendations from inactive users. Furthermore, Xian et al.~\cite{xian2019reinforcement} proposed a reinforcement learning approach for recommendation, where explanations are delivered by a policy-guided path reasoning on a KG. Wang et al.~\cite{wang2019explainable} suggested recommendation explanations by showing items shared features and mining association rules. Lastly, Huan et al.~\cite{huang2019explainable} proposed a method to provide path-based explanation through user-rating interactions incorporating dynamic user preferences. Our recommendation method brings entities and relationships in the same latent space, and presents embedding-based explanations, which have been proven to be effective~\cite{ma2019jointly}. 

\section{Proposed approach}

\begin{figure}[t]
    \centering
    \includegraphics[scale=0.35,trim={0mm 8mm 0mm 8mm}]{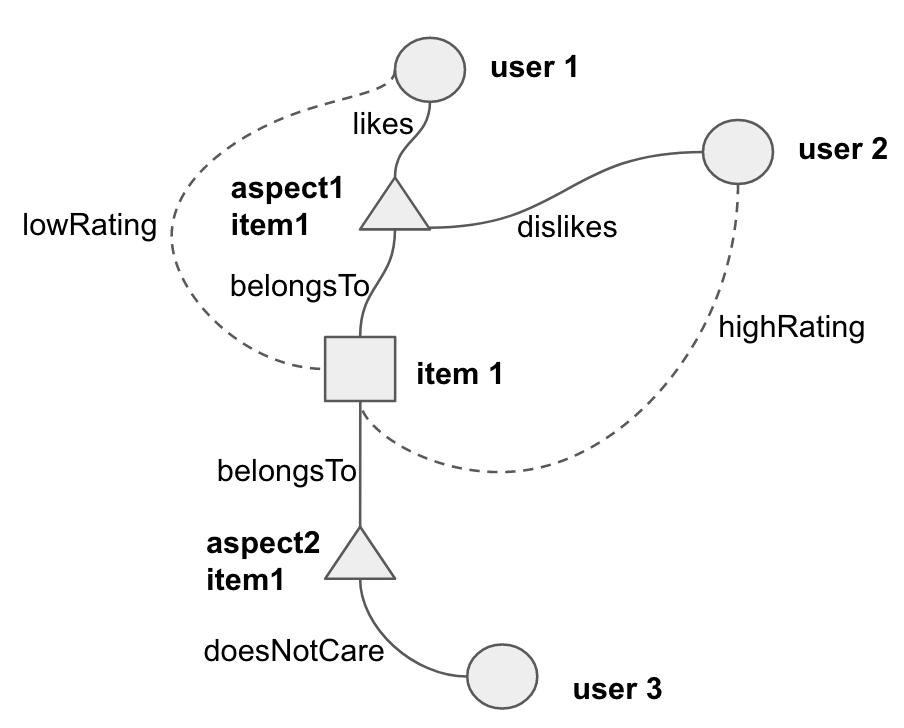}
    \caption{Example of ratings and aspect-based opinions graph. Users may \textit{like}, \textit{dislike} or \textit{doesNotCare} about item aspects, or may assign \textit{highRating} or \textit{lowRating} values to items. An aspect \textit{belongsTo} an item. Circle, triangle and square nodes represent users, aspects and items, respectively. Dotted lines correspond to rating relations, while solid lines are aspect-based opinion relations.}
    \label{graph_fig}
\end{figure}

\subsection{Knowledge graph representation}
Our KG representation model is depicted in Figure \ref{graph_fig}. Let $G = (N, R, E)$ be a graph where \textit{N} are nodes representing users, items and item aspects; \textit{R} is the set of possible relations between such entities: \textit{like} and \textit{dislike} for user-aspect relations, \textit{belongsTo} for aspect-item relations, and \textit{highRating} and \textit{lowRating} for user-to-item relations; and \textit{E} represents the graph edges, whose elements are semantic triple $<s,r,d>$ --source, relation, and destination--, where $s,d \in N$ and $r \in R$.  
More specifically, to represent ratings given by users to items as triples $<s,r,d>$, we use two types of relations: \textit{highRating} and \textit{lowRating}. If a rating is lower or equal to 3, then it is transformed into a \textit{lowRating} relation; otherwise, it is represented as \textit{highRating} relation. An aspect-based opinion, by contrast, is represented as two triples $<s,r,d>$: one that relates the user and the aspect, and other that related the aspect and the item. The former relation can be a \textit{likes}, \textit{doesNotCare} or \textit{dislikes} relation depending on the polarity (positive, neutral or negative) of the opinion given to the aspect. This polarity is computed through the opinoin mining technique proposed by Hernandez-Rubio et al.~\cite{hernandez2019comparative}. If the extracted polarity is $0$, it means the user \textit{doesnotCare} about the aspect for that particular item; if the polarity is greater than $0$, the user \textit{likes} the item's aspect, and finally, if the polarity is lower than $0$, the \textit{dislikes} the item's aspect. The latter relation is always \textit{belongsTo}, stating that the aspect is associated to the item. 
An illustrative KG example is shown in Figure \ref{graph_fig}. user1 likes aspect1, which belongs to item1, despite the fact that the item receives a low rating from the user. user2 dislikes the above aspect of item1, but assigns a high rating to the item. Finally, user3 does not care about aspect2 of item1, and does not rate the item.

\subsection{Knowledge graph embedding recommendations}
We used PyTorch BigGraph (PBG), a graph embedding learning framework released by Facebook AI~\cite{lerer2019pytorch}, to generate user, item, and aspect embeddings from the proposed KG. Then, we used the learned embeddings to make personalized recommendations. To explain the PBG representation learning process we define $E$ as a set of edges represented as triplets $<s,r,d>$ that connect source (\textit{s}) and destination (\textit{d}) nodes using a relation (\textit{r}). For each triplet, PBG model have parameters $(\theta_s, \theta_r, \theta_d)$ representing source node ($s$), relation ($r$) and destination node ($d$), and a function $f$ whose purpose is to maximize the cases where the triple $e$ belongs to $E$ and seek to minimize the other cases. To achieve this, $f$ obtain a latent representation learned for $s$, $d$ and $r$: $f(\theta_s, \theta_r, \theta_d) = sim(g_s(\theta_s, \theta_r), g_s(\theta_d, \theta_r)$. Where the transformation function $g$ encodes the relation between node and relation $r$ vectors, and the similarity function ($sim$) calculates a similarity between both transformed vectors. PBG combines similarity ($sim$) and transformation ($g$) functions to learn embeddings for source node, destination node, and relations, by minimizing the following loss function: $L = \sum_{e \in E} \sum_{e' \in S'_e} max(f(e) - f(e') +\lambda , 0) $. Where $S'_e$ is a set of edges obtained using negative sampling, which samples half of them from corrupting one node from training edges and the other half from triples randomly created, and $\lambda$ is a margin hyperparameter. The combination of the similarity and the scoring function depends on the type of relations being modeled. In our case, we use complEX \cite{trouillon2016complex} which uses dot product as similarity function, and $Re(<s,\bar{d}>)$ as transformation function, since we empirically obtained that it was the best combination in terms of recommendation performance compared to transE~\cite{bordes2013translating}, transR~\cite{lin2015learning}, RESCAL~\cite{nickel2011three}, Distmult~\cite{yang2014embedding}, among others. Finally, to make the recommendation, as user and item embeddings are in the same latent space, we look for the item embeddings closest to the user embedding using cosine similarity. 

\vspace{-1mm}

\section{Datasets}
To evaluate the proposed method, we used the Amazon Reviews datasets\footnote{https://jmcauley.ucsd.edu/data/amazon/} (AMZ) with reviews and ratings for Movies \& TV (MVT), Videogames (VGM), and Cellphones (CPH). For this dataset, we considered only users with more than $10$ ratings. We also considered other domains --Restaurants (RST), Hotels (HTL), and Beauty\&Spa (BTY)-- using the Yelp Challenge dataset\footnote{https://www.yelp.com/dataset} (YLP), which, as well as the Amazon dataset, contains both rating and reviews. Then, concerning the aspect opinions dataset\footnote{http://ir.ii.uam.es/aspects/}, this information complements the rating information since it includes users' polarity with aspects of items given ratings depending on the content of the reviews' text.
Statistics from datasets are shown in Table \ref{table-datasets}, it can be seen that the number of ratings varies from 25 thousand to more than one million and a half interactions. It can be seen that Yelp datasets for HTL and BTY domains have fewer interactions compared to others. For aspect-based relations, the number of interactions varies from 13 thousand interactions to more than 2 million interactions for the MVT domain. It can be seen that, in general, users are more active for rating interactions more than aspect opinions, except for the MVT domain. 

\vspace{1mm}

\begin{table}[!htb]
\small
\caption{Datasets statistics}
\vspace{-3mm}
\label{table-datasets}
\centering
{
\begin{tabular}{ccccccccc} 
    \toprule
    {Dataset} & {Domain} & {Users} & {Items} & {Ratings}  & {Aspect} & {Rating} \\
    &&&&&opinions&sparsity
    \\ \hline
    
    \multirow{3}{*}{\makecell{AMZ}} & {MVT} & 40,058 & 131,638 & 1,349,351 & 2,132,927 & $2.4 \cdot 10^{-4}$   \\
    {} & {VGM} & 6,686 & 25,795 & 575,136 & 209,755 & $8.3 \cdot 10^{-4}$    \\ 
    {} & {CPH} & 7,237 & 47,183 & 464,684 & 85,930 &  $3.4 \cdot 10^{-4}$  \\ 
    
    \hline
    
    \multirow{3}{*}{\makecell{YLP}} & {RST} & 36,471 & 4,503 & 791,865 & 547,892 & $9.6 \cdot 10^{-4}$  \\
    {} & {HTL} & 4,148 &  284 & 25,170 & 16,935 & $4.2 \cdot 10^{-3}$ \\ 
    {} & {BTY} & 4,270 & 764 & 27,885 & 13,217 &  $1.7 \cdot 10^{-3}$  \\ 
    
    \bottomrule
    
    \end{tabular}}
\end{table}

\begin{table*}[]
\small
\caption{Examples of recommendation explanations for the MTV and RST domains, where top 30 similar users have positive or negative opinions about aspects of the recommended items.}
\label{explanation_examples}
\vspace{-3mm}
\begin{tabular}{ccccc}
\toprule
    Recommended item & Similar 30 users' aspect opinions & Score & Sample review  \\
    \hline 
     Avengers & \multicolumn{1}{m{4cm}}{characters (15 \faThumbsUp , 5 \faThumbsDown) \newline 
                                        acting (10 \faThumbsUp , 3 \faThumbsDown) \newline
                                        picture (2 \faThumbsUp , 10 \faThumbsDown)} 
              & 4.3  
              &  \multicolumn{1}{m{10cm}|}{I enjoy all of the Marvel movies. It's not as good as the later 3 that follows, Ultron, infinity wars and Endgame but it's still cool to see a lot of the Marvel \textbf{characters} come together for the first time. The \textbf{graphics} are great for the year and the \textbf{acting} is very good as well.}  \\
    
    \hline 
     Get\&Go Burrito & \multicolumn{1}{m{4cm}}{service (18 \faThumbsUp , 4 \faThumbsDown) \newline 
                                        price (10 \faThumbsUp , 14 \faThumbsDown) \newline
                                        atmosphere (17 \faThumbsUp , 13 \faThumbsDown)
                                        } 
              & 3.5  
              &  \multicolumn{1}{m{10cm}}{Know 2 things - this place isn't \textbf{cheap} and you shouldn't eat for a couple of days before you go!  Know that we've gotten that out of the way, let me tell you - t his \textbf{place} is great!  The \textbf{service} was fabulous.}  \\

\bottomrule
\end{tabular}
\end{table*}

\vspace{-4mm}

\section{Experiments}

\subsection{Evaluated methods}
We report experimental results from the following methods:

\begin{itemize}
    
    \item \textbf{Non-personalized recommendation methods:} we consider random rating (RDM) and most popular item (POP) recommendation methods. 
    
    \item \textbf{Matrix Factorization optimized with alternate least squares (MF):} as a traditional baseline, we chose matrix factorization optimized with alternate least-squares~\cite{koren2009matrix} using 200 latent factors, which yield the best results among other traditional baselines. 
    
    \item \textbf{Graph-based recommendation baseline (KGAT): } as a graph-based recommendation state-of-the-art baseline, we consider Wang et al.'s KG Attention Network recommendation framework (KGAT)~\cite{wang2019kgat} . 
    
    \item \textbf{Graph embeddings from ratings (GER):} this is a version of our proposed model that considers only rating-based relations between users and items.
    
    \item \textbf{Graph embeddings from aspect-based opinions (GEA):} this is a version of our proposed model that considers only \textit{likes}, \textit{dislikes}, and \textit{doesNotCare} relations between users and item aspects, and \textit{belongsTo} relation between item aspects and items. 
    
    \item \textbf{Graph embeddings from ratings and aspect-based opinions (GERA):} this is our proposed model that considers both aspect and rating-based relations.
    
\end{itemize}

Content-based filtering, user-based collaborative filtering~\cite{schafer2007collaborative}, item-based collaborative filtering~\cite{schafer2007collaborative}, matrix factorization using alternated least-squares~\cite{koren2009matrix}, and matrix factorization using Bayesian personalized ranking~\cite{rendle2012bpr} were also evaluated, performing worse. We do not report their results due to lack of space.

\vspace{-2mm}

\subsection{Experimental setup}
We tested a range of parameters, obtaining best performance results for embeddings of dimension 400 and learning rate 0.01. In the evaluation, we used the OrdRec methodology~\cite{koren2013collaborative}, where a method is evaluated on the test ratings at the user level. We employ a 5-fold cross-validation strategy, and compute the F1@k (10,20,30) metric. Other metrics such as nDCG@k(10,20,30), recall@k(10,20,30), and precision@k(10,20,30) were also calculated, but they are not reported herein due to lack of space; moreover, they did not give additional insights to the analysis we present next. 

\vspace{-2mm}

\subsection{Achieved results}
The results in Tables \ref{amazon_results_table} and \ref{yelp_results_table} indicate that in most of the domains our method (GERA) outperforms the baselines in terms of F1@k (10,20, 30). For a top-N recommendation task, GERA outperforms state-of-the-art traditional and graph-based recommendation methods. However, for the CPH domain, in terms of F1@10, MF achieves the highest value (0.915), whereas, for F1@20 and F1@30, the best-performing method was the aspect-based opinion graph (GEA) version of our proposal, with values of 0.935 and 0.939, respectively. The worse results of GERA in the CPH domain can be explained because, as shown in Table \ref{table-datasets} this domain does not have enough aspect opinions to improve the results by combining them with ratings in a KG. 

\begin{table}[]
        \small
        \captionof{table}{Evaluation results on Amazon dataset. Values in bold are the best ones for each domain and metric. * indicates statistical significance best result by multiple t-test using Bomferroni correction.}
        \label{amazon_results_table}
\vspace{-3mm}

        \begin{tabular}{cccccc}
        \toprule
         Dataset & Domain & Model & F1@10 & F1@20 & F1@30  \\ \hline
    
         \multirow{6}{*}{}& {}  & {RDM}  & {.801}  & {.812} & {.813}   \\
         && POP & .033 & .031 & .031    \\
         && MF & .905 & .914 & .917   \\
         & MVT &KGAT & .924 & .938 & .942 \\
         &&GER & .921 & .936 & .940   \\
         &&GEA & .915 & .932 & .936   \\
         &&GERA & \textbf{.927*} & \textbf{.941*} & \textbf{.945*}  \\

        \cline{2-6}
        
         && RDM & .800 & .807 & .806  \\
         && POP & .045 & .045 & .045    \\
         && MF & .909 & .914 & .915 \\ 
         AMZ & VGM &KGAT & .791  & .903 & .920  \\
         && GER & .791 & .901 & .920  \\
         &&GEA  & .781 & .898 & .918 \\
         &&GERA & \textbf{.934*} & \textbf{.943*} & \textbf{.945*}      \\

        \cline{2-6}
        
        && RDM  & .819 & .819 & .823   \\
        && POP & .043  & .043 & .043   \\
        && MF & \textbf{.915*} & .916 & .917             \\ 
        & CPH &KGAT & .888 & .929  & .933 \\
        && GER & .890 & .930 & .934 \\
        && GEA & .898 & \textbf{.935*} & \textbf{.939*} \\
        && GERA & .880 & .928 & .933   \\

        \bottomrule
        
        \end{tabular}

\vspace{7mm}

        \small
        \captionof{table}{Evaluation results on Yelp dataset. Values in bold are the best ones for each domain and metric. * indicates statistical significance best result by multiple t-test using Bomferroni correction.}
        \label{yelp_results_table}
\vspace{-3mm}

        \begin{tabular}{cccccc}
        
        \toprule
        
        Dataset & Domain & Model & F1@10 & F1@20 & F1@30  \\ \hline
        
        \multirow{3}{*}{} &  &  RDM & .647 & .646 & .650 
          \\
         && POP & .107 & .107 & .107   \\
         && MF & .827 & .829 & .830    \\ 
         & RST &KGAT & .913 & .932 & .938 \\
         && GER & .914  & .931  & .936 \\
         && GEA & .927 & .940 & .944 \\
         &&GERA & \textbf{.933*} & \textbf{.944*} & \textbf{.947*}             \\ 
         
         \cline{2-6}
        
         && RDM & .592 & .589 & .571    \\ 
         && POP &.468 & .468 & .468   \\
         && MF & .816 & .816 & .816    \\
         YLP & HTL &KGAT & .986 & .986 & .986\\
         &&GER & .983 & .983 & .983 \\
         &&GEA & .985 & .985 & .985 \\
         &&GERA & \textbf{.987*} & \textbf{.987*} & \textbf{.987*}    \\

        \cline{2-6}
        
         && RDM & .650 & .634 & .622     \\
         && POP & .263 & .263 & .263   \\
         && MF & .846 & .848 & .850    \\
         & BTY &KGAT & .984 & .984 & .984\\
         &&GER& .983 & .984 & .984 \\
         &&GEA& .983 & .983 & .983  \\
         &&GERA & \textbf{.987*} & \textbf{.988*} & \textbf{.988*}    \\ 
         
        \bottomrule
        \end{tabular}
\end{table}

\vspace{-2mm}

\section{Recommendation Explanations}
In this section, we present a simple technique to generate explanations about recommendations provided by our method. To obtain interpretable recommendations at aspect level, we first recommend 30 (a chosen cut-off) items whose embeddings are closer to the user's embedding. From them, we search for the 30 most similar users, and obtain the opinions these users gave to the aspects of the recommended items. In Table \ref{explanation_examples}, we show examples of generated aspect-level explainable recommendations. In Table \ref{table-intepretability}, we report \textit{Coverage} --percentage of items with aspect opinion-based explanations--, \textit{lk/other} --the proportion of \textit{likes} over \textit{dislikes} and \textit{doesNotCare}) relations--, \textit{\#aspects} --the average number of unique aspects--, and \textit{asp/item} --the average number of aspects per item.

\begin{table}[H]
\caption{Statistics of the recommendation explanation technique for each domain. Results in bold are the best ones for each metric.}
\setlength{\tabcolsep}{0.45em} 
{\renewcommand{\arraystretch}{0.95}
\label{table-intepretability}
\centering
\vspace{-3mm}
{
\small
\begin{tabular}{ccccccccc} \toprule
    {Dataset} & {Domain} & {Coverage} & {lk/other} & {\#aspects} & {asp/item}    \\ \hline
    
    \multirow{3}{*}{\makecell{AMZ}} & {MVT} & 47.90\% & .663 & 10.30 & .716   \\
    {} & {VGM} & \textbf{90.23}\% & .584  & 18.52  & .684     \\ 
    {} & {CPH} & 70.38\% & .592 & 12.27  & .581  \\ 
    
    \hline
    
    \multirow{3}{*}{\makecell{YLP}} & {RST} & 11.11\% & .508 & 6.22 & 1.87   \\
    {} & {HTL} & 21.64\% & \textbf{.796} & \textbf{22.34} & \textbf{3.44} \\ 
    {} & {BTY} & 20.02\% & .653 & 17.55 & 2.88  \\ 
    
    \bottomrule
    
    \end{tabular}}}
\end{table}

\vspace{-2mm}    

The results in Table \ref{table-intepretability} indicate that the proposed method has better coverage in the VGM domain with 90.34\% coverage; this shows that of the 30 recommended items, the proportion mentioned above may have interpretation at the aspect level. Regarding the balance of likes with respect to other interactions, the proposed method in the HTL domain has the highest ratio (0.796). This means that, in general, the aspects of hotels are valued higher than in other domains. The method explanations in that domain cover a greater variety of aspects, with an average of 22.34 unique aspects per recommendation. More evidence of this is given by the high number of aspects per item (3.44). However, despite the diversity in explanations, the method in the HTL domain has the third worst coverage, followed by RST and BTY. Comparing the results on the two datasets, it can be seen that in terms of coverage, in general, AMZ surpasses YLP.

\section{Conclusions} 
In this paper, we have empirically shown that latent representations of users and items extracted from a KG by means of rating-based embeddings can be exploited to effectively outperform the performance of graph-based and traditional recommendation approaches. Moreover, we have seen that richer embeddings integrating ratings and aspect-based opinions in a KG allows surpassing a state-of-the-art neural network recommendation model. 

Another advantage of the proposed method is the interpretabilty of its recommendations, since  users, items, and aspects are modeled in a common latent space. Differently to previous approaches that generate explanations at rating level or according to graph paths that require domain knowledge, 
our approach provides aspect-level explanations extracted from item reviews.

Our work thus shows that to improve state-the-art graph-based recommendations, it is not always necessary to change the recommendation algorithm, but preferably seek new ways of incorporating existing information like ratings and aspect-based opinions into the used KG.



\bibliographystyle{ACM-Reference-Format}
\bibliography{sample-base}


\end{document}